\documentclass[11pt]{article} 
\usepackage{fullpage}
\usepackage{tcolorbox}
\usepackage{wrapfig}
\tcbuselibrary{breakable}
\usepackage{times}
\usepackage[utf8]{inputenc} 
\usepackage[T1]{fontenc}    
\usepackage{hyperref}
\usepackage{url}            
\usepackage{booktabs}       
\usepackage{amsfonts}       
\usepackage{nicefrac}       
\usepackage{microtype}      
\usepackage{tablefootnote}
\usepackage{makecell}

\usepackage{amsfonts}       
\usepackage{amssymb}
\usepackage{amsmath}
\usepackage{amsthm}
\usepackage{mathtools}
\usepackage{multicol}
\usepackage{multirow}

\theoremstyle{definition}

\usepackage{color}

\usepackage{array}
\usepackage{algorithm}
\usepackage{algorithmic}

\usepackage[shortlabels]{enumitem}

\usepackage[symbol]{footmisc}

\usepackage{pifont}
\usepackage{wrapfig}

\usepackage{subcaption}
\usepackage{caption}
\usepackage[normalem]{ulem}
\usepackage{tikz}
\usepackage{color}
\usepackage{array}
\usepackage{algorithm}
\usepackage{algorithmic}

\newcounter{gaocomm}

\definecolor{blue-violet}{rgb}{0.00,0.75,0.90}
\definecolor{mygreen}{rgb}{0.0, 0.5, 0.0}
\definecolor{awesome}{rgb}{1.0, 0.13, 0.32}
\definecolor{bostonuniversityred}{rgb}{0.8, 0.0, 0.0}

\usepackage{soul}
\usepackage{color}
\usepackage[normalem]{ulem}
\newcounter{ToDo1}
\newcounter{guocomm}
\newcounter{Note1}
\definecolor{blue-violet1}{rgb}{0.54, 0.17, 0.89}
\definecolor{mygreen}{rgb}{0.0, 0.5, 0.0}
\definecolor{awesome}{rgb}{1.0, 0.13, 0.32}
\definecolor{wsuacdred}{rgb}{0.93, 0.0, 0.2}
\definecolor{wsucrimson}{rgb}{0.6, 0.0, 0.2}

\newcommand{\guorm}[1]{\ignorespaces}

\usepackage[utf8]{inputenc}

\title{\textbf{Machine Learning-Based Prediction of Key Genes Correlated to the Subretinal Lesion Severity in a Mouse Model of Age-Related Macular Degeneration}}

\author{Kuan Yan \footnote{Discipline of Business Analytics, Business School, The University of Sydney, Camperdown, NSW 2006, Australia
(\text{kuan.yan, dai.shi, dmytro.matsypura, junbin.gao@sydney.edu.au}).}
\and Yue Zeng\footnote{Macula Research Group, Save Sight Institute, Faculty of Medicine and Health, The University of Sydney, Camperdown, NSW 2006, Australia. (\text{yue.zeng, ting.zhang, mark.gillies,ling.zhu\}@sydney.edu.au@sydney.edu.au})}\footnote{Department of Ophthalmology, The First Affiliated Hospital of Zhejiang University, Hangzhou, Zhejiang Province, China.} \footnotemark[2]\footnotemark[3]
\and Dai Shi \footnotemark[1] 
\and Ting Zhang \footnotemark[2]
\and Dmytro Matsypura \footnotemark[1]
\and Mark C. Gillies \footnotemark[2]
\and Ling Zhu \footnotemark[2] \,\,\footnote{Co-corresponding autors: {junbin.gao,lingzhu}@sydney.edu.au}\footnotemark[4]
\and  Junbin Gao \footnotemark[1]\,\,\footnotemark[4]
}
\date{}

\begin{document}

\maketitle

\begin{abstract}
Age-related macular degeneration (AMD) is a major cause of blindness in older adults, severely affecting vision and quality of life. Despite advances in understanding AMD, the molecular factors driving the severity of subretinal scarring (fibrosis) remain elusive, hampering the development of effective therapies. This study introduces a machine learning-based framework to predict key genes that are strongly correlated with lesion severity and to identify potential therapeutic targets to prevent subretinal fibrosis in AMD. Using an original RNA sequencing (RNA-seq) dataset from the diseased retinas of JR5558 mice, we developed a novel and specific feature engineering technique, including pathway-based dimensionality reduction and gene-based feature expansion, to enhance prediction accuracy. Two iterative experiments were conducted by leveraging Ridge and ElasticNet regression models to assess biological relevance and gene impact. The results highlight the biological significance of several key genes and demonstrate the framework's effectiveness in identifying novel therapeutic targets. The key findings provide valuable insights for advancing drug discovery efforts and improving treatment strategies for AMD, with the potential to enhance patient outcomes by targeting the underlying genetic mechanisms of subretinal lesion development.
\end{abstract}

\section{Introduction}
Neovascular age-related macular degeneration (nAMD) is the leading cause of blindness in individuals aged 50 and older \cite{blindness2021vision}. While anti-vascular endothelial growth factor (VEGF) therapies have been the gold standard for treating nAMD, long-term studies reveal that up to 70\% of patients treated with anti-VEGF drugs develop subretinal fibrosis within 10 years, resulting in severe visual loss \cite{gillies2020ten}\cite{bloch2013subfoveal}. Currently, no FDA-approved treatments exist for subretinal fibrosis, making the identification of novel genetic targets and biological pathways critical for improving visual outcomes in nAMD. 

The spontaneous JR5558 mouse model \cite{won2011mouse}, developed at the Jackson Laboratory, offers a valuable tool for studying the progression of nAMD. These mice develop subretinal fibrovascular lesions, visible as yellow mounds in fundus photographs, starting at 4 weeks and expanding until 12 weeks of age. A critical angio-fibrotic switch occurs at around 8 weeks, making the model particularly suited for examining both early neovascular changes and late-stage fibrosis \cite{nagai2014spontaneous, hasegawa2014characterization, linder2024vivo}. Therapeutic targets for subretinal fibrosis have been validated in this model \cite{rossato2020fibrotic}. However, traditional methods of studying nAMD, which rely on observational and statistical techniques, often fall short due to the complexity and vastness of genetic data. These conventional approaches lack the precision and scalability required to identify specific genes that drive disease progression, underscoring the need for novel strategies that can address these challenges.

Machine Learning (ML) offers significant potential in the field of genomics, with its ability to analyze vast datasets and uncover intricate patterns that may not be apparent through conventional analysis \cite{bostanci2023machine}\cite{chandrasekhar2023enhancing}. By applying ML models to RNA sequencing (RNA-seq) data, researchers can predict lesion severity and identify key genes associated with subretinal fibrosis. Consequently, using ML to focus on disease severity in the JR5558 mouse model can further aid in high-throughput screening for novel therapeutic targets in subretinal fibrosis, thereby advancing drug discovery efforts. However, RNA-seq datasets can vary significantly across different applications, often facing challenges such as limited sample sizes and high dimensionality. Effectively preprocessing these datasets, selecting the most suitable models and designing appropriate prediction tasks have become critical challenges.

To address the aforementioned issues, we investigate a novel problem in this paper: how to utilize ML models to predict lesion severity and identify potential therapeutic gene targets in subretinal fibrosis related to AMD using RNA-seq data from mice. First, we collected and organized comprehensive RNA-seq data from the retinas of JR5558 mice. We then preprocessed this original RNA-seq dataset by reducing dimensionality and expanding features before training the ML models. We employed Ridge and ElasticNet regression models for training and prediction. To verify the effectiveness of our proposed framework and further analyze the biological impact, we conducted two sets of iterative experiments on biological correlation and gene impact measurement.

The main contributions of our research can be summarized as follows:
\begin{itemize}
\item We collected and provided an original and comprehensive RNA-seq dataset from the retinas of JR5558 mice, facilitating further research in subretinal fibrosis.
\item We applied ML models, specifically Ridge and ElasticNet regression, to predict lesion severity and identify key genes associated with subretinal fibrosis, thereby enhancing precision in genetic analysis.
\item We tackled challenges related to limited sample sizes and high dimensionality in RNA-seq data, improving data preprocessing and model selection strategies in transcriptomic research.
\item We designed and conducted iterative experiments based on the original datasets we collected and produced, verifying the effectiveness and excellent performance of our proposed framework through biological impact analysis in two dimensions: biological correlation and gene impact measurement.
\item Our approach identifies potential therapeutic gene targets, offering new insights into transcriptomic influences on subretinal fibrosis and advancing drug discovery efforts.
\end{itemize}

The remainder of this paper is structured as follows. Section~\ref{Sec:2} presents a literature review of machine learning in biomedical research, RNA-seq data in machine learning and transcriptomic studies on disease severity and subretinal fibrosis. Section~\ref{Sec:3} illustrates our proposed framework and methodology. Section~\ref{Sec:4} details the dataset description, presents and discusses the experimental results and biological impact and Section~\ref{Sec:5} summarizes the paper’s conclusions.

\section{Related Work}\label{Sec:2}
\subsection{Machine Learning in Biomedical Research}
In recent years, the integration of ML into biomedical research has revolutionized the field, offering novel insights and predictive capabilities that were previously unattainable. As the volume of biomedical data continues to grow exponentially, ML provides essential methods for analyzing intricate datasets, discovering patterns and making informed predictions.

Traditional statistical methods, such as t-tests and ANOVA, have been instrumental in biomedical research but come with inherent limitations. These methods often require assumptions about the data and can struggle with the high dimensionality and complexity typically for biological datasets. Furthermore, traditional approaches may not always capture subtle patterns and interactions within the data, leading to potential oversights. In contrast, ML offers significant advantages in handling large and complex datasets. ML algorithms can uncover intricate patterns and relationships without the need for predefined models, enhancing the ability to make accurate predictions and discoveries. This flexibility is particularly beneficial in multi-omics studies, where the complexity of the data demands more robust and adaptive analytical techniques.

ML techniques have significantly advanced biomedical research, enhancing our ability to address various critical aspects such as predicting disease outcomes, identifying biomarkers and uncovering therapeutic targets. In the realm of predicting disease outcomes, Khan et al. \cite{khan2023novel} and Bhatt et al. \cite{bhatt2023effective} focused on cardiovascular diseases, with Khan’s employing random forest (RF) algorithms and Bhatt’s utilizing a multilayer perceptron with cross-validation, both significantly enhancing early diagnosis precision. Arumugam et al. \cite{arumugam2023multiple} optimized a decision tree model to predict heart disease in diabetes patients, improving diagnostic accuracy and efficiency. Similarly, Islam et al. \cite{islam2023chronic} evaluated various ML models for chronic kidney disease prediction, determining that the XGBoost classifier was the most effective, thereby supporting timely intervention and treatment planning. Transitioning to the identification of biomarkers, Zhang et al. \cite{zhang2021machine} employed feature selection methods alongside various ML techniques, including support vector machines (SVMs), to boost diagnostic accuracy and aid in the development of targeted treatments. Mi et al. \cite{mi2021permutation} introduced PermFIT, a permutation-based technique using RFs and SVMs, to identify crucial biomarkers for complex diseases. This method isolated key genetic markers, enhancing the understanding and diagnosis of intricate medical conditions. In the quest to uncover therapeutic targets, Pun et al. \cite{pun2023ai} utilized deep learning models to analyze large-scale omics data, identifying novel therapeutic targets for complex diseases by uncovering critical molecular interactions and pathways. Rafique et al. \cite{rafique2021machine} applied ensemble learning methods, such as RFs and gradient boosting machines, to predict patient responses to various cancer treatments. By analyzing clinical and molecular data, they aimed to improve the accuracy of therapeutic response predictions, ultimately enhancing patient outcomes in oncology.

With continuous advancements in this field, ML has demonstrated significant advantages in handling large-scale biological datasets and uncovering meaningful patterns. Consequently, it has become a valuable tool for processing and analyzing RNA-seq data.

\subsection{RNA Sequencing Data in Machine Learning}
RNA-seq has significantly advanced biomedical research and transcriptomics by offering a comprehensive view of gene expression. Unlike traditional profiling methods, RNA-seq quantifies transcript levels across the entire genome, providing insights into gene regulation and cellular functions \cite{slovin2021single}. This detailed analysis enables the identification of transcriptomic activity patterns associated with disease processes, which are crucial for understanding complex biological systems. Additionally, it aids in discovering potential biomarkers and therapeutic targets \cite{andrews2021tutorial}. In ML applications, the ability of RNA-seq data to capture the full range of gene expression makes it an invaluable resource for developing predictive models. These models can elucidate the transcriptomic factors that influence disease severity and progression.

RNA-seq data have been instrumental in uncovering critical insights across a range of conditions, ultimately contributing to improved diagnosis, treatment and understanding of complex diseases. For instance, in oncology, RNA-seq has been used to classify cancer subtypes that predict cancer progression and treatment response \cite{yu2020rna}. Additionally, in neurodegenerative diseases such as Alzheimer’s, RNA-seq has revealed critical interactions between gene pairs that contribute to disease mechanisms, offering potential targets for intervention \cite{chen2019machine}. Furthermore, in cardiovascular research, RNA-seq has provided insights into genes associated with heart failure and atrial fibrillation, aiding in the development of predictive models that enhance disease prediction and support precision medicine \cite{venkat2023investigating}.

Building on these advancements, ML models play a crucial role in leveraging RNA-seq data for disease research. Supervised learning models, such as SVMs and RFs, have been effectively utilized to identify biomarkers and predict treatment responses. Gupta et al. \cite{gupta2021identifying} deployed RF and SVM to analyze RNA-seq datasets for identifying and validating novel transcript biomarkers associated with hepatocellular carcinoma, focusing on improving early detection and diagnosis. Meanwhile, unsupervised learning models like hierarchical clustering have uncovered novel gene expression patterns in cancer research, offering valuable insights into underlying mechanisms. Lee et al. \cite{lee2020deep} proposed an approach that uses similarity-based hierarchical clustering to accurately analyze complex patient data and identify distinct patterns that correlate with disease progression, thereby enhancing the prediction of pathological stages in papillary renal cell carcinoma.

To enhance the efficacy of ML models applied to RNA-seq datasets, it is essential to conduct several preprocessing steps on the raw data. These steps typically include quality control measures to remove low-quality reads, serving as an initial data denoising process. In addition, one can deploy feature normalization to account for variations in sequencing depth and filtering to eliminate uninformative genes. Moreover, feature extraction methods, such as gene expression quantification and dimensionality reduction techniques such as principal component analysis (PCA) \cite{chen2020robust}, are crucial for identifying relevant features that capture the most informative aspects of the RNA-seq data.

\subsection{Molecular Studies on the Pathogenesis of Subretinal Fibrosis}
Subretinal fibrosis is a hallmark of advanced neovascular age-related macular degeneration (nAMD) and is closely associated with poor visual outcomes \cite{tenbrock2022subretinal}. Although anti-VEGF therapies have revolutionized the treatment of nAMD by suppressing neovascularization, their long-term effectiveness is limited. A substantial proportion of patients, despite initial responsiveness to anti-VEGF therapies, develop subretinal fibrosis within a decade, leading to irreversible vision loss \cite{khachigian2023emerging}. Fibrosis, characterized by the excessive accumulation of extracellular matrix (ECM) components beneath the retina, results in tissue scarring and visual impairment \cite{mallone2021understanding}. This underscores the need for a deeper understanding of the molecular pathways contributing to fibrotic processes, as current treatment strategies are insufficient in halting or reversing fibrosis progression.

Recent research has uncovered critical molecular factors that influence the severity of subretinal fibrosis. Specifically, mutations in genes responsible for ECM remodeling have been implicated as key drivers of fibrosis in nAMD \cite{shughoury2022molecular}. Collagen and fibronectin, structural proteins of the ECM, are essential for maintaining tissue integrity and facilitating repair processes \cite{nita2014age}. However, genetic mutations in these proteins can promote aberrant ECM deposition, exacerbating fibrotic tissue development. Furthermore, the interplay between genetic predispositions and environmental factors, such as oxidative stress and chronic inflammation, accelerates the fibrotic response in the subretinal space \cite{kauppinen2016inflammation}. Inflammatory signaling pathways, such as those mediated by cytokines and chemokines, are known to amplify the fibrotic process by enhancing the recruitment of fibroblasts and the deposition of ECM components, leading to retinal scarring. This interaction between genetic and environmental factors highlights the complexity of fibrosis and the need for multifaceted therapeutic approaches.

\section{Methodology}\label{Sec:3}
\subsection{Framework Overview}
\begin{figure}[H]
\centerline{\scalebox{0.8}{\includegraphics[width=1.0\linewidth]{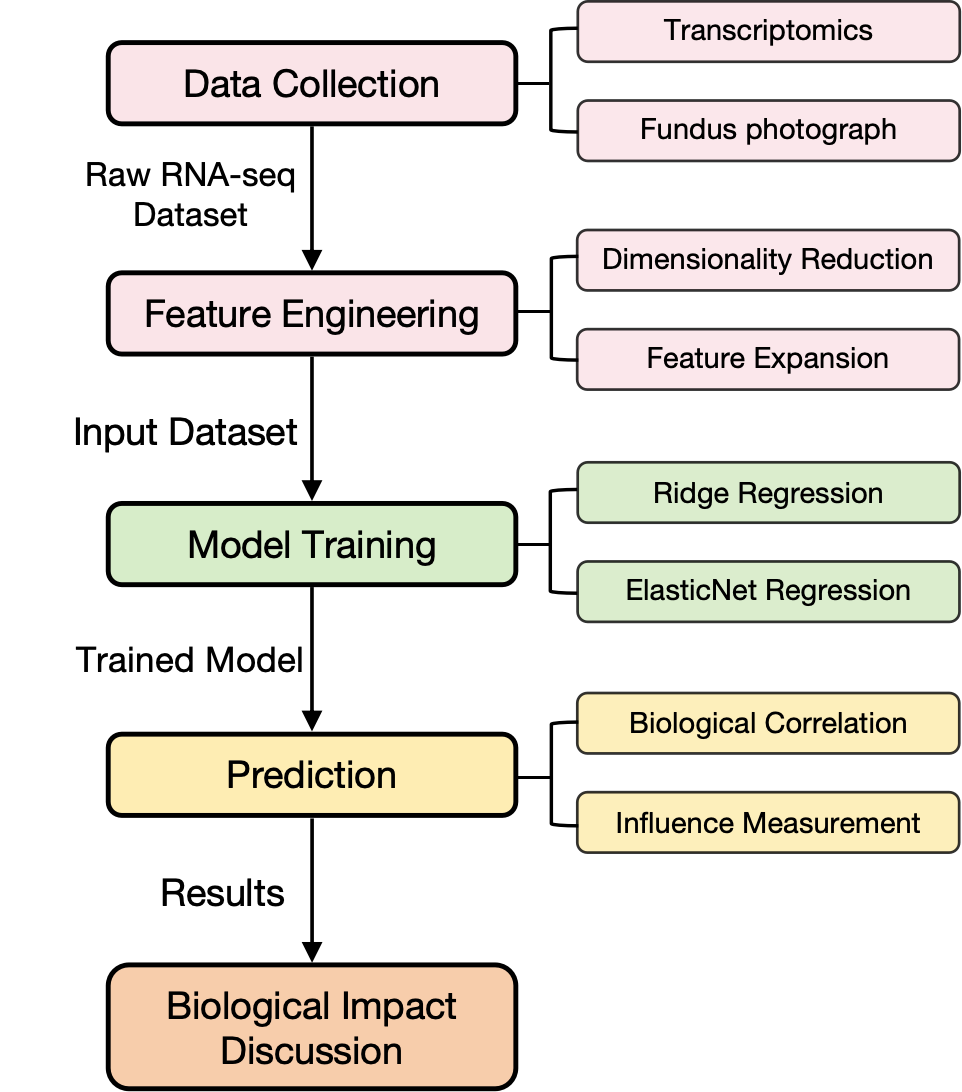}}}
    \caption{The overall framework of our study.}
    \label{fig:01_framework}
\end{figure}

Our study, as depicted in Figure~\ref{fig:01_framework}, employs a comprehensive ML-based framework designed to predict lesion severity and identify key gene targets associated with the disease using RNA-seq data from the JR5558 mouse model. This approach aims to deepen our understanding of the genetic factors influencing disease progression and to uncover potential therapeutic targets.

In the initial phase of our research, we collected RNA-seq data from the retinas of 23 JR5558 mice. This data was meticulously correlated with lesion severity scores obtained from fundus photographs of these mice, where severity was quantified by measuring subretinal lesion size. This process provides us with a foundational raw RNA-seq dataset, crucial for subsequent analyses. Next, in the feature engineering stage, we tackled the challenge posed by the limited sample size and the extensive number of features. This was achieved through dimensionality reduction, specifically by organising the data according to group genes into different molecular pathways. This method allows us to focus on the most pertinent gene groups. Once these influential gene groups were identified, we expanded the dataset to concentrate on individual genes within these groups, enhancing the dataset’s utility for further analysis. Following this, taking the refined dataset as input, we proceeded to the training and prediction phase using two ML models: Ridge regression and ElasticNet regression. These models were chosen for their ability to handle complex data and provide accurate predictions. Then, based on our trained models, we conduct two iterative experiments: biological correlation and influence measurement. The objective of the first experiment is to identify genes whose expression is most strongly associated with the severity of subretinal lesions in AMD, while the second experiment aims to identify target genes that, when modified, could significantly alter lesion severity, potentially revealing new treatment targets for subretinal fibrosis in AMD. Finally, we delved into a comprehensive discussion of the biological impact of our findings. This analysis is pivotal in exploring and identifying potential therapeutic targets, which may offer new avenues for the treatment of subretinal fibrosis. 

By integrating these stages, our framework provides a robust and systematic approach to understanding the genetic influences on lesion severity, ultimately contributing valuable insights for future therapeutic interventions.

\subsection{Animals}
JR5558 mice were purchased from Jackson Laboratory (B6.Cg-$Crb1^{rd8}$ $Jak3^{m1J}$/Boc: JAX stock\# 005558), with a genetic background of C57BL/6J. All animals were housed in the pathogen-free environment of the Animal Research Facility on a 12-light/dark cycle. The experimental procedures were conducted in accordance with the ARVO Statement for the Use of Animals in Ophthalmic and Vision Research (\url{http://www.arvo.org/}). This study was approved by the Animal Ethics Committee of the University of Sydney (Project number: 2021/2013).

\subsection{Data Collection}
\begin{figure}[t]
\centerline{\scalebox{0.8}{\includegraphics[width=0.95\linewidth]{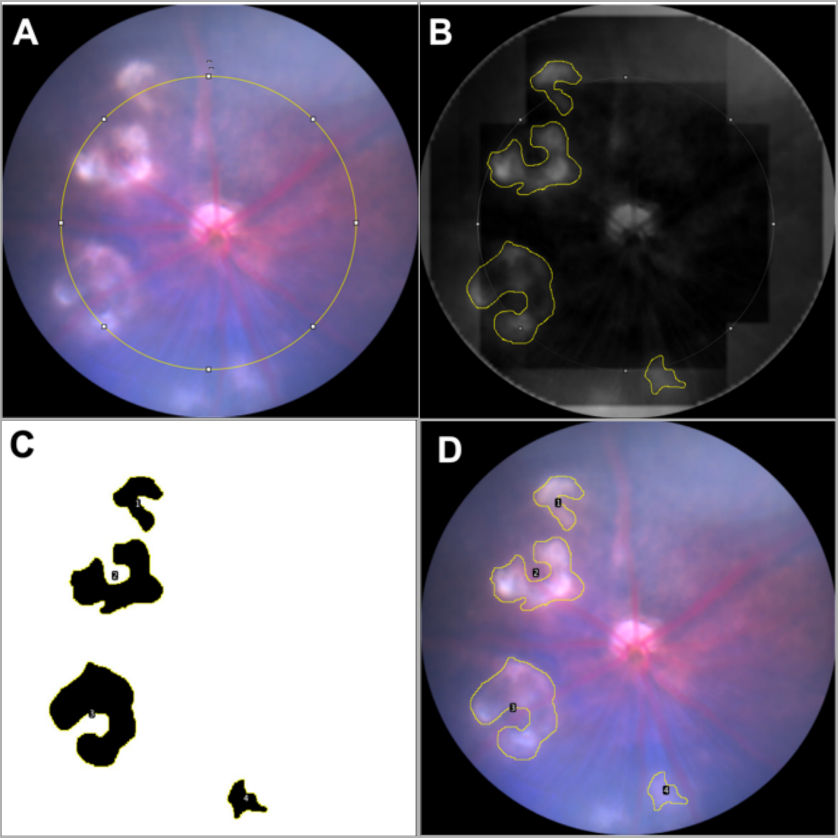}}}
    \caption{Key steps of the quantification method.}
    \bigskip 
    A: Draw a circle with a radius of 283 $\mu$$m$ centered on the optic nerve. B: Convert the image to an 8-bit grayscale, remove the background and mark with freehand tools. C: Adjust the threshold to precisely select the lesion area. D: Verify in the original image that all lesions have been selected.
    \label{fig:02_fundus}
\end{figure}

The study utilizes bulk RNA-seq data from the retinas of JR5888 mice, a well-established model for studying nAMD. Total RNA from twenty-three retinas of 8-week-old male JR5558 mice was extracted using GenEluteTM Single Cell RNA Purification Kit (Sigma Aldrich, RNB300). The library preparation, quality control and sequencing were commercially contracted to Novogene (\url{https://www.novogene.com/}). The RNA-seq dataset includes expression levels of 56,748 genes, quantified as Fragments Per Kilobase of transcript per Million mapped reads (FPKM). The FPKM values provide a normalized measure of gene expression, accounting for gene length and sequencing depth, allowing for accurate comparisons across samples. Out of the 56,748 genes, 24,888 have corresponding Entrez IDs, which were used for subsequent pathway analysis. In addition to the RNA-seq data, fundus photographs were also taken to analyze the disease severity. In brief, mice were anesthetised by intraperitoneal injection of ketamine (48 $mg/kg$, Troy Laboratories, Australia) and medetomidine (0.6 $mg/kg$, Troy Laboratories, Australia). Mice pupils were dilated with 0.5\% Tropicamide. Fundus photographs were performed with MICRON IV Retinal Imaging Microscope (Phoenix Technology Group, USA) with optic nerve locating at the center of the image. The total area of subretinal lesions on fundus images was then independently quantified by two investigators using FIJI ImageJ software ( \url{https://imagej.net/software/fiji/downloads}, National Institutes of Health, USA). Figure~\ref{fig:02_fundus} illustrates the key steps of the quantification method. An imageJ macro was developed accordingly to ensure consistent output. Of note, lesions that either totally or partially fell into the circle with a radius of 283 $\mu$$m$ from the optic nerve were included in the analysis. Lesion severity was quantified based on the percentage of subretinal lesion area observed in fundus photographs of the mouse retina. The lesion area serves as the primary outcome variable for predicting the severity of fibrosis in this study.

\subsection{Feature Engineering}

\subsubsection{Dimensionality Reduction}
Dimensionality reduction is a crucial technique in the analysis of RNA-seq data, particularly when investigating the influence of genes on disease severity. In relation to our research on the JR5558 mouse model, dimensionality reduction serves to strike a balance between model performance and computational efficiency. High-dimensional data can lead to overfitting, where the model becomes too tailored to the training data of small size, resulting in poor generalisation to unseen data. Conversely, reducing the dimensionality excessively may lead to the loss of vital biological information necessary for understanding disease progression. 

Several methods are available for dimensionality reduction, including feature extraction techniques like PCA and feature selection methods such as Least Absolute Shrinkage and Selection Operator (LASSO) and RF. For our study, feature selection is particularly advantageous, as it retains the interpretability of the model. This is essential for identifying potential therapeutic targets, as we aim to elucidate the relationship between individual genes and the severity of subretinal fibrosis.

Incorporating domain knowledge is essential for effective feature selection from RNA-seq data, as it helps identify relevant genes that influence disease severity. Without this understanding, ML models may miss important non-linear relationships, potentially excluding critical genes. To maintain robust model performance, feature selection should be performed solely on the training dataset to prevent data leakage, which could skew performance metrics. Additionally, ensuring consistent variable types across training and testing datasets is crucial for achieving reliable predictions and generalizability.

In our study, we face a challenge common to many biological datasets: a limited number of samples from JR5558 mice, yet a vast array of gene features for each sample. To address this, we developed an approach to effectively reduce the dimensionality of the RNA-seq data by grouping genes into canonical molecular pathways. A web scraper was developed in R to extract detailed biological pathway information related to Mus musculus from the KEGG database (\url{https://www.genome.jp/kegg-bin/show_organism?menu_type=pathway_maps&org=mmu}). The data extracted included pathway maps, which were organized into a data frame for subsequent dimensional reduction. This step was crucial to managing the high dimensionality of the RNA-seq data by focusing on relevant pathways rather than individual genes, thus reducing noise and improving the model’s ability to identify significant correlations. By grouping gene features and averaging expression values within each group, we were able to reduce the dataset’s dimensionality from over 24,888 features to 343, thereby streamlining the dataset for more effective prediction and analysis.

\subsubsection{Feature Expansion}
Feature expansion is a technique used to enhance the predictive capabilities of a model by increasing the number of relevant features. It involves transforming existing data to uncover additional insights that may not be immediately apparent. This method can be particularly useful in complex biological datasets, allowing for a more comprehensive analysis by incorporating diverse aspects of the data.

Pathway-based dimensionality reduction allows for effective predictive analysis on gene-group datasets with reduced dimensionality. Once we identify the key gene groups, we can expand their original gene features to enhance training and prediction focused on these crucial genes. Specifically, genes from the identified key biological pathways are extracted and their FPKM values from each sample are used as features for the second round of data processing. This gene-based feature expansion allows the model to incorporate detailed expression information for genes that are potentially involved in fibrosis, enhancing its predictive power.

\section{Experiments}\label{Sec:4}
\subsection{Dataset Description}
The dataset for this study comprises bulk RNA-seq data from the retinas of JR5888 mice and corresponding lesion area measurements. The RNA-seq data included expression levels for 56,748 genes, with 24,888 genes associated with Entrez IDs. The lesion area data were derived from fundus photographs, where the extent of subretinal fibrosis was quantified as a percentage of the total retinal area. This dataset was divided into training and validation sets to evaluate the performance of the ML models in predicting key genes associated with disease severity.

\subsection{Prediction and Results}
\begin{figure}[H]
    \centerline{\scalebox{0.8}{\includegraphics[width=1.0\linewidth]{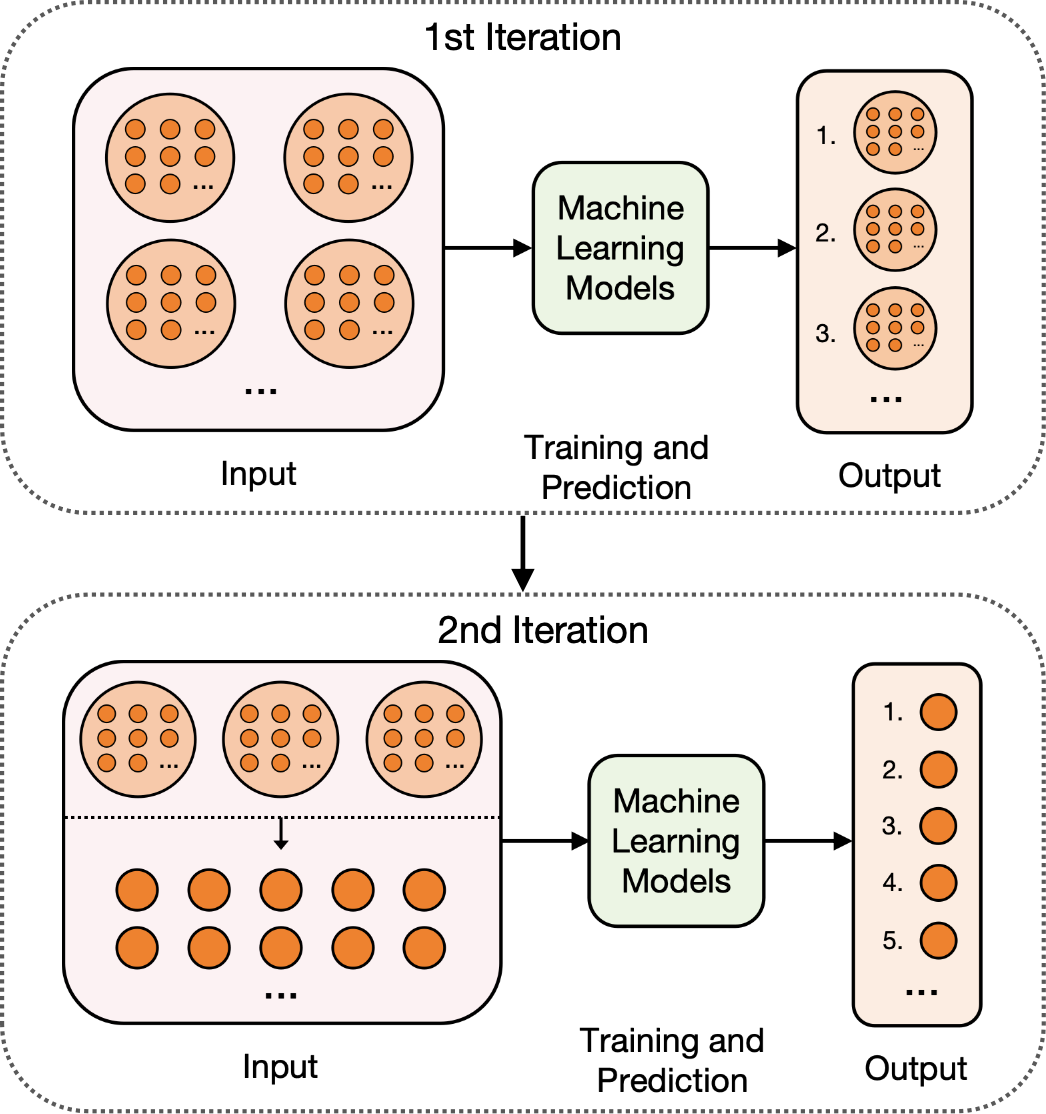}}}
    \caption{The logic of our iterative experiments.}
    \label{fig:03_logic}
\end{figure}

In this study, we conducted two prediction tasks: 1. biological significance and 2. influence measurement, each involving two rounds of iterative experiments. Fig.~\ref{fig:03_logic} illustrates the logic of our experimental approach. In the first round, we employed a dataset with gene groups as features, utilizing dimensionality reduction through a gene pathway-based method to perform predictive tasks. Based on these initial results, we identified the top-performing gene groups as candidates for further analysis. In the second round, we expanded our focus by using the expression values of the original genes within these selected candidates. This allowed us to conduct more detailed prediction tasks and perform biological analysis targeting individual genes, thereby enhancing the depth and accuracy of our findings.

In this study, we conducted two prediction tasks: 1. biological significance and 2. influence measurement, each involving two rounds of iterative experiments. Figure~\ref{fig:03_logic} illustrates the logic of our experimental approach. In the first round, we employed a dataset with gene groups as features, utilizing dimensionality reduction through a gene pathway-based method to perform predictive tasks. Based on these initial results, we identified the top-performing gene groups as candidates for further analysis. In the second round, we expanded our focus by using the expression values of the original genes within these selected candidates. This allows us to conduct more detailed prediction tasks and perform biological analysis targeting individual genes, thereby enhancing the depth and accuracy of our findings.

\begin{table*}[htbp]
\caption{Top 7 most important genes selected by our ML models in biological correlation experiments with the greatest association to the disease.}
\begin{center}
\scalebox{0.9}{
\begin{tabular}{c|c|c}
\toprule
\textbf{Ranking}  & \textbf{Influential Gene Entrez ID} & \textbf{Gene Name and Description}\\ 
\midrule
1   &16177	&interleukin 1 receptor, type I  \\
2	&18796	&phospholipase C, beta 2 \\
3	&12259	&complement component 1, q subcomponent, alpha polypeptide  \\
4	&320302	&glycosyltransferase 28 domain containing 2 \\
5	&15212	&hexosaminidase B \\
6	&18018	&nuclear factor of activated T cells, cytoplasmic, calcineurin dependent 1 \\
7	&20848	&signal transducer and activator of transcription 3 \\
\bottomrule
\end{tabular}}
\label{table1}
\end{center}
\end{table*}

\subsubsection{Biological Correlation}
The objective of the first task is to predict the genes most strongly correlated with the severity of disease progression in AMD.

In the first round of experiments, we used Ridge regression and ElasticNet regression models to train and predict the gene group dataset following dimensionality reduction, identifying the overlapping most influential gene groups from both models. We identified nine recurring candidates from the top ten most influential gene groups produced by both models. In the second round of experiments, we expanded the gene expression features within these influential gene groups to further identify the most significant genes. The seven selected most influential genes, which have the greatest association with the disease in biological correlation experiments, are listed in Table \ref{table1}.

\subsubsection{Influence Measurement}
The purpose of the second task is to identify target genes that, when manipulated, could significantly modulate the severity of subretinal fibrosis, potentially leading to more effective therapeutic strategies.

In the first round, we adjusted each gene group feature by decreasing its expression value to 50\% and increasing it to 200\%, respectively. We then applied the trained model to predict lesion severity based on these modified gene group expressions and recorded the changes in predicted severity. We calculated the differences in lesion severity before and after the adjustments to each gene group’s expression. This allows us to identify the gene groups whose modifications resulted in the most significant changes in lesion severity. Here, we counted and ranked the changes in lesion severity into two categories: aggravation and alleviation. Therefore, we identified the most significant gene groups that either worsened or alleviated the disease as candidates. By the end of the first round of experiments, we obtained two rankings for the gene groups: 50\% manipulation that alleviated the disease, and 200\% manipulation that caused the disease to worsen.

\begin{table*}[htbp]
\caption{Top 10 genes selected by our ML models for their significant impact on disease improvement when gene expression values were reduced by 50\%.}
\begin{center}
\begin{tabular}{c|c|c}
\toprule
\textbf{Ranking}  & \textbf{Influential Gene Entrez ID} & \textbf{Gene Name and Description}\\ 
\midrule
1	&66357	&oligosaccharyltransferase complex subunit (non-catalytic)\\
2	&26416	&mitogen-activated protein kinase 14\\
3	&12260	&complement component 1, q subcomponent, beta polypeptide \\
4	&16179	&interleukin-1 receptor-associated kinase 1\\
5	&320302	&glycosyltransferase 28 domain containing 2\\
6	&12262	&complement component 1, q subcomponent, C chain\\
7	&224530	&acetyl-Coenzyme A acetyltransferase 3\\
8	&14676	&guanine nucleotide binding protein, alpha 15\\
9	&16176	&interleukin 1 beta\\
10	&12503	&CD247 antigen\\
\bottomrule
\end{tabular}
\label{table2}
\end{center}
\end{table*}

\begin{table*}[htbp]
\caption{Top 10 genes selected by our ML models for their significant impact on disease exacerbation when gene expression values were increased to 200\%.}
\begin{center}
\begin{tabular}{c|c|c}
\toprule
\textbf{Ranking}  & \textbf{Influential Gene Entrez ID} & \textbf{Gene Name and Description}\\ 
\midrule
1	&18798	&phospholipase C, beta 4\\
2	&12260	&complement component 1, q subcomponent, beta polypeptide\\
3	&12259	&complement component 1, q subcomponent, alpha polypeptide\\
4	&12322	&calcium/calmodulin-dependent protein kinase II alpha\\
5	&12262	&complement component 1, q subcomponent, C chain\\
6	&19091	&protein kinase, cGMP-dependent, type I\\
7	&106759	&toll-like receptor adaptor molecule 1\\
8	&20293	&chemokine (C-C motif) ligand 12\\
9	&224530	&acetyl-Coenzyme A acetyltransferase 3\\
10	&12789	&cyclic nucleotide gated channel alpha 2\\
\bottomrule
\end{tabular}
\label{table3}
\end{center}
\end{table*}

In the second round of experiments, we performed feature expansion on the two ranked gene group lists from the first round and obtained two corresponding datasets with individual gene expression values as features. For each dataset, we performed the same manipulations as in the previous round and counted the rankings that caused the same disease impact. This time, we identified the most influential individual genes. For example, for the gene groups that resulted in 50\% manipulation causing the disease alleviation in the previous round, we continued to apply the same 50\% manipulation after expanding the gene dataset and ranked the genes that had the same impact on the disease, which was improving. This process allowed us to identify the genes that most significantly affected lesion severity. Tables \ref{table2} and \ref{table3} have shown the final results of the influence measurement experiments, respectively.

\subsection{Biological Impact Discussion}
The integration of ML models with RNA-seq data offers a robust approach for identifying transcriptomic factors linked to disease severity in AMD. This study specifically aimed at predicting genes associated with the severity of subretinal lesions in a mouse model of AMD. By employing dimensionality reduction and feature expansion techniques, our model successfully identified genes within specific pathways that their related proteins and signaling pathways are likely contributors to subretinal fibrosis that may be considered as therapeutic targets. These findings enhance our understanding of the molecular mechanisms underlying subretinal fibrosis and present new opportunities for therapeutic interventions.

It is remarkable that two different tasks with different algorithms identified several common key genes/proteins or targeting same pathway. We first compared the top 10 target genes in table II and table III and identified two proteins is positively correlated to the lesion severity. They are Complement component 1, q subcomponent (C1q) and Acetyl-Coenzyme A acetyltransferase 3 (ACAT3). We further compared the results from task 1 and task 2 and further two more common proteins has been identified. They are Phospholipase C (PLC), and Glycosyltransferase 28. All four common proteins were strongly implicated in both the progression of subretinal fibrosis and as potential therapeutic targets (Tables \ref{table1}, \ref{table2} and \ref{table3}). Each of these proteins operates within distinct yet interconnected biological processes, highlighting a complex interaction between inflammation, immune responses, cellular signaling and metabolic regulation that drives fibrotic changes in the retina.

The C1q proteins are key components of the classical complement pathway which initiates immune surveillance, inflammation and the clearance of apoptotic cells \cite{cho2019emerging} \cite{galvan2012c1q}. Dysregulation of this pathway, particularly sustained activation of C1q, has been closely linked to the progression of AMD \cite{ma2022association}. Ongoing C1q activation in subretinal fibrosis fuels chronic inflammation, promoting tissue damage and extracellular matrix remodeling—both hallmarks of fibrosis. The accumulation of matrix components in the subretinal space results in tissue thickening and scarring, which contribute to irreversible loss of vision in advanced AMD. It is worth noting that we also identified several inflammation-related molecules, including Interleukin 1 receptor, type I (IL1R1, Table \ref{table1}), Signal transducer and activator of transcription 3 (STAT3, Table \ref{table1}), Interleukin-1 receptor-associated kinase 1 (IRAK1, Table \ref{table2}), Interleukin 1 beta (IL1B, Table \ref{table2}), Toll-like receptor adaptor molecule 1 (TICAM1, Table \ref{table3}) and Chemokine (C-C motif) ligand 12 (CCL12, Table \ref{table3}). This suggests a strong positive correlation between the activation of inflammatory pathways and the severity of the subretinal lesion. 

PLC is critical for intracellular signal transduction. This signaling pathway is also essential for regulating cellular responses to inflammatory stimuli, proliferation and apoptosis \cite{wu2023roles}. PLC may intensify complement activation in subretinal fibrosis by amplifying pro-inflammatory signals \cite{zhu2018role}. This interaction between PLC and the complement system may create a self-perpetuating cycle of inflammation that drives tissue remodeling and fibrosis in the retina.

Acetyl-Coenzyme A acetyltransferase (ACAT) plays a key role in acetyl-CoA metabolism by influencing cholesterol and production of ketone bodies \cite{ma2023hepatic}. Its involvement in lipid metabolism is particularly relevant to retinal diseases, where metabolic dysregulation often exacerbates inflammation and tissue damage \cite{ana2023precision}. Altered ACAT activity could disrupt lipid homeostasis, promoting cellular stress and inflammation in retinal cells, which may accelerate the development of subretinal fibrosis by further stimulating inflammatory and fibrotic processes.

Glycosyltransferase is an enzyme responsible for glycosylation, the addition of sugar moieties to proteins and lipids. This modification impacts protein folding, stability and interactions, all of which are crucial for maintaining cellular function \cite{nagare2021glycosyltransferases}. Aberrant glycosylation has been associated with fibrosis across various tissues \cite{loaeza2021overview}. In subretinal fibrosis, Glycosyltransferase may influence key protein modifications involved in inflammation and extracellular matrix formation, potentially exacerbating tissue scarring and the progression of fibrosis.

Together, these four proteins, C1q, PLC, ACAT3 and Glycosyltransferase, likely form an interconnected network that sustains chronic inflammation and metabolic dysfunction in the retina. Their combined activity promotes extracellular matrix deposition and tissue remodeling, driving the progression of subretinal fibrosis. Understanding their precise roles and interactions could provide critical insights into therapeutic strategies for AMD. Targeting these proteins or their signaling pathways may offer effective ways to reduce inflammation, slow fibrotic changes and prevent vision loss in patients with AMD.

These results underscore the potential of combining ML with molecular imaging techniques to enhance our understanding of fibrotic diseases and improve patient outcomes. Future research could explore the application of this approach to other fibrotic conditions and assess its potential for personalising treatment strategies based on individual genetic profiles.

\section{Conclusions}\label{Sec:5}
In this paper, we have presented a comprehensive approach to addressing the challenges of predicting lesion severity in nAMD using an ML-based framework. We introduced a unique RNA-seq dataset derived from the JR5558 mouse model, which provides a valuable resource for further research in subretinal fibrosis. We successfully addressed issues of limited sample sizes and high dimensionality typical of RNA-seq data by employing dimensionality reduction and feature expansion techniques. Utilizing Ridge and ElasticNet regression models, our iterative experiments confirmed the effectiveness of our framework, highlighting its potential in identifying critical genetic targets linked to subretinal fibrosis.

The insights gained from our study have substantial implications for genetic research and therapeutic development. We offer new avenues for drug discovery and improved treatment strategies for nAMD by pinpointing key gene targets, ultimately aiming to enhance patient care. Our research underscores the importance of integrating advanced ML techniques in genomic studies, paving the way for future investigations that further connect genetic findings with clinical applications.

\bibliographystyle{plain}
\bibliography{ref_copy}

\end{document}